\documentclass{aastex}

\newcommand{\Rsun}{\ensuremath{R_{\odot}}}
\newcommand{\Msun}{\ensuremath{M_{\odot}}}

\def \ie{{\sl i.e.}}
\def \eg{{\sl e.g.}}
\def \viz{{\sl viz.}}

\usepackage{graphicx,natbib,subfigure}

\title{Collective Properties of X-ray Binary Populations of Galaxies III.
  \\ The Low-mass X-ray Binary Luminosity Function}

\author{H. Bhadkamkar\altaffilmark{1, 2} and P. Ghosh\altaffilmark{1} }

\altaffiltext{1}{Department of Astronomy \& Astrophysics, Tata
        Institute of Fundamental Research, Mumbai 400 005, India}

\altaffiltext{2}{Astronomy \& Astrophysics, Raman Research Institute,
        Bengaluru, 560080}

\begin{abstract}
Continuing our exploration of the collective properties of
low-mass X-ray binaries (LMXBs) in the stellar fields (\ie, 
outside globular clusters) of normal galaxies, we compute in
this paper (Paper III in the series) the expected X-ray 
luminosity function (XLF) of LMXBs, starting from the results
of the previous paper in the series (Paper II). For this, we    
follow the evolution of the pre-LMXB population (described
in Paper II) through Roche-lobe contact and the consequent LMXB
turn-on, LMXB evolution through the accretion phase, and the 
eventual conclusion of accretion and LMXB turn-off. We treat 
separately two classes of LMXB evolution, the first (a) being 
close systems whose initial orbital periods are below the 
bifurcation period, wherein the companion is on the main 
sequence when Roche-lobe contact occurs, the subsequent
evolution is driven by angular-momentum loss from the system,
and the second (b) being wider systems whose initial orbital 
periods are above the bifurcation period, wherein the companion 
is on the giant branch when Roche-lobe contact occurs, and the 
evolution of these systems is driven by the nuclear evolution 
of the companion. We obtain model luminosity profiles $L(t)$
for individual LMXBs of both classes, showing that they are in 
general agreement with those in previous works in the subject. 
We point out that the basic features of the luminosity profile 
in the angular-momentum-loss driven case can be well-understood
from scaling laws first pointed out by Patterson (1984) in a 
related context, which we call \emph{Patterson scaling}. We
then compute the LMXB XLF by ``folding in'' the inputs for the 
pre-LMXB collective properties and formation rates from Paper II
with the above luminosity profiles. Because of the long timescale 
($\sim 10^9$ yr) on which LMXBs evolve, this computation becomes 
more involved, as one needs to keep track of the evolution of 
the star-formation rate (SFR) on the same timescale, and we use
star-formation histories (SFH) given by canonical models. 
We compare the observed LMXB XLF with our computed one, keeping in
mind that we have included only neutron-star systems in this work,
so that there would be an unaccounted-for population of black-hole
binaries at the high-luminosity end. We show that a qualitative 
similarity already exists between the two, and we discuss the role
of the \emph{giant fraction}, \ie, that fraction of all LMXBs
which harbors a low-mass giant companion, on the shape of the XLF 
at the high-luminosity end.       
\end{abstract}


\begin{document}
\maketitle

\section{Introduction}
\label{intro}

This is Paper III in our series of papers exploring the basic physics
underlying the \emph{collective} properties of accretion powered
X-ray binaries. In Paper II, we described our study of the constraints
on the formation of the so-called pre-low-mass X-ray binaries 
(pre-LMXBs), which is a relatively rare process, and our calculation
of the collective properties (\ie, distributions of their essential 
parameters) and the rates of formation of these pre-LMXBs. In this 
paper, we use the results of Paper II to compute the LMXB X-ray 
luminosity function (XLF).

Our procedure here is to first follow the evolution of the 
pre-LMXB population through Roche-lobe contact and consequent LMXB
turn-on, LMXB evolution through the accretion phase, and eventual 
conclusion of accretion and LMXB turn-off. To this end, we  
calculate the X-ray luminosity evolution of individual model systems
in our computational scheme, and so construct luminosity profiles,
$L(t)$, for these systems. We treat separately the two well-known 
categories of LMXB evolution. The first is that for close systems
whose initial orbital periods are below the so-called \emph{bifurcation 
period}, wherein the companion is on the main sequence when 
Roche-lobe contact occurs and mass transfer starts. Evolution of
such systems is driven by angular-momentum loss from the systems 
through gravitational radiation and magnetic braking, and their 
orbits shrink as mass transfer proceeds. The second is that for
wider systems whose initial orbital periods are above the bifurcation
period, wherein the companion has finished its main-sequence 
evolution and is on the giant branch when Roche-lobe contact occurs 
and mass transfer starts. Evolution of these systems is driven by
nuclear evolution and expansion of the companion, and their orbits
expand as mass transfer proceeds. We obtain model luminosity profiles
for both categories, and show that they are in general agreement
with those in previous works in the subject. We point out that the 
basic features of the luminosity profile in the angular-momentum-loss 
driven category can be accounted for by scaling laws first 
pointed out by \citet{patterson} in a related context, which we 
call \emph{Patterson scaling}.         
  
We obtain the LMXB XLF by combining the above inputs. Because of the 
long timecasle ($\sim 10^9$ yr) on which LMXBs evolve, the 
computation here becomes more involved, as one needs to keep 
track of the evolution of the star-formation rate (SFR) on the same
timescale, since the rate of formation of primordial binaries,
which evolve into LMXBs, is determined by the SFR. Accordingly,
we construct a scheme for ``folding in'' the inputs from the 
pre-LMXB collective properties and formation rates, together with 
the star-formation history (SFH), \ie, the time-variation profile 
of the SFR, as given by canonical models in wide use. We thus 
obtain the theoretical LMXB XLF and study its properties and 
their variations with some parameters characterising the primordial 
binary distribution the supernova process.

We compare the observed LMXB XLF with our computed one, keeping in
mind that we have included only neutron-star systems in this work,
so that there would be an unaccounted-for population of black-hole
binaries at the high-luminosity end. We show that a qualitative 
similarity already exists between the two, and we discuss the role
of the \emph{giant fraction}, \ie, the fraction of all LMXB systems
harboring a low-mass giant companion, on the shape of the XLF at
the high-luminosity end.

The rest of the paper is organized as follows. In Sec.\ref{evolsec},
we describe the essential aspects of LMXB evolution that we need
for our purposes here. We give the prescriptions for angular 
momnetum loss by graviational radiation and magnetic braking, and
we describe the two categories of LMXB evolution in terms of the
bifurcation period. We obtain the luminosity profile for  
angular-momentum-loss driven evolution, and introduce and discuss
the Patterson scaling (1984) which accounts for it. We then obtain
the luminosity profile for evolution driven by nuclear evolution
of the companion. In Sec.\ref{scheme}, we describe our scheme for
computing the XLF through the device of convolving the inputs on
pre-LMXBs from Paper II with these luminosity profiles, together 
with the SFH for taking into account the simultaneous evolution
of the SFH. We discuss the properties of our computed XLF,
exploring the effects of some primordial-binary parameters and
supernova characteristics on it. In Sec.\ref{discuss}, we 
compare our computed XLF with observations, and discuss the 
results. Our conclusions are presented in Sec.\ref{conclude}.

\section{LMXB Evolution}
\label{evolsec}

A pre-LMXB system is initially detached. Mass transfer and 
accretion onto the neutron star is not possible in this state and
becomes so only when the system evolves, its orbit shrinks
due to loss of angular momentum, and Roche-lobe contact is achieved. 
The orbital angular momentum of the binary is given by

\begin{equation}
J^2 = G \,\frac{M_c^2M_{NS}^2}{M_t} a
\label{eqn:jorb}
\end{equation}

Here $M_c$ and $M_{NS}$ are the masses of the companion and the neutron star
respectively and $M_t = M_{NS}+M_c$ is the total mass of the system. $a$ is
the binary separation. 

We describe binary orbital evolution due to angular momentum loss by 
differentiating Eq.(\ref{eqn:jorb}) with respect to time. We first
note that $\dot{M}_t = \dot{M}_{NS} + \dot{M}_c$.
Since the companion is the mass donor, we can write $\dot{M}_c = 
-\dot{M}$ where  $\dot{M}$ is the mass-transfer rate.
Allowing for non-conservative mass transfer in general, we write 
$\dot{M}_{NS} = \gamma \dot{M}$, \ie, a fraction $\gamma$
of the mass transferred by the donor is accreted onto the neutron star and
the rest of the mass is lost from the system, reducing its total mass
at a rate $\dot{M}_t = -(1-\gamma)\dot{M}$. This mass loss implies
an appropriate loss of angular momentum from the system.
 It is customary to represent this loss of angular momentum
in terms of the specific angular momentum of the system 
\citep{belc08, pfahl03}, expressing the result as
$\dot{J}/J = \mu \, \dot{M}_t/M_t$. Here, $\mu$ is
typically of order unity, its exact value depending on the binary
parameters as well as the evolutionary state of the donor. By inserting
the expressions for the rates of change of various quantities as described
above, the differential form of eqn. \ref{eqn:jorb} can be rewritten as:

\begin{equation}
\left(\frac{\dot{J}}{J}\right)_{NML} = \frac{\dot{M}}{M_c}
\left[\gamma q - 1 + \frac{(1-\gamma)(1+2\mu)}{2}\frac{q}{1+q} \right]
+ \frac{\dot{a}}{2a}
\label{eqn:jorbdiff}
\end{equation}

Here, $(\dot{J}/J)_{NML}$ is the rate of loss of angular momentum from
the system due to processes \emph{other than} the above mass loss from 
the system, and $q=M_c/M_{NS}$ is the mass ratio. We now consider 
such processes of angular momentum loss.

\subsection{Mechanisms of angular momentum loss}
\label{jloss}

Two mechanisms have been widely studied in the literature that cause the 
angular momentum loss in the pre-LMXB systems. These are gravitational
radiation and magnetic braking. We now briefly review these two 
mechanisms and discuss their effects on the binary parameters.

It is well-known that two point masses orbiting around each other
would emit gravitational waves according to the general theory of relativity
\citep{peters63, peters65}, and the rate of evolution of the orbital 
separation (as also the eccentricity in case of eccentric orbits) due to 
this effect is well-known. It was shown by \citet{faulkner} that this 
effect can be important in close binary systems like CVs and LMXBs.
The timescale for the rate of angular momentum loss is comparable to 
typical LMXB evolutionary timescales, which makes it an essential 
effect in LMXB evolution. The rate of loss of the angular
momentum due to this effect is given for a circular orbit  by:

\begin{equation}
\left(\frac{\dot{J}}{J}\right)_{GW} = -0.831 \frac{M_cM_{NS}M_t}{a^4}
\label{eqn:gw}
\end{equation}

\emph{We shall use the following system of units throughout this paper, 
unless stated otherwise. All the masses and the radii/distances are given 
in terms of the solar mass and the solar radius respectively, and time
is in the units of Gyr. X-ray luminosity of the LMXBs will be given 
in units of $10^{36}$ erg/s and will be denotes as $L_{36}$.}

We observe the strong dependence on the orbital separation in 
Eq.(\ref{eqn:gw}), which makes gravitational radiation completely 
ineffective at large separations. However, since pre-LMXB orbits are 
already very compact due to large contractions during the CE phase, 
this mechanism can be, and often is, the dominant one determining the
orbit shrinkage that converts pre-LMXBs into LMXBs.

Magnetic braking was suggested as a possible  mechanism for angular 
momentum loss by \citet{verbunt} in order to explain the observed high 
rates of the mass transfer in some LMXB systems. This mechanism 
depends upon the mass loss due to a magnetic wind from a tidally
locked companion, working as follows. The magnetic field of the
companion makes the matter in its coronal region co-rotate with the
star, so that this matter has a large specific angular momentum, and
consequently even a moderate mass-loss rate can lead to a large rate
of angular momentum loss. Since the companion is tidally locked in 
pre-LMXBs and LMXBs, this loss of angular momentum will 
ultimately come from the orbital angular momentum of the binary,
making it shrink. The rate of loss of angular momentum due to 
magnetic braking is given by

\begin{equation}
\left(\frac{\dot{J}}{J}\right)_{MB} = -57.2 \,\eta\,
\frac{M_t^2}{M_{NS}} \frac{R_c^4}{a^5}
\label{eqn:mb}
\end{equation}

Here $R_c$ is the radius of the companion and $\eta$ is a parameter of 
order unity which can be empirically adjusted to fine-tune the strength of the
magnetic braking. We note here that the magnetic braking is effective only 
when the star has a considerable magnetic activity. It has been suggested 
that magnetic activity would suddenly decrease drastically when the 
star becomes fully convective. This happens when the mass of the star falls 
below $\sim 0.3\Msun$. Thus the magnetic braking is expected to turn off 
when the companion mass falls below this limit, and this is believed to 
be the reason for the observed period gap in CV systems.

Some authors have suggested lower strengths of magnetic braking than 
in the original formulation (see, \eg, \citet{sluys}), implemented by 
either (a) reducing the braking by a fixed fraction ($\eta = 0.25$,
say), or (b) using different functional
forms in two different regions of angular velocity. It was also suggested
by \citet{podsiadlowski} that this reduction in strength may depend
upon the mass of the convective envelope.
The calculational scheme for collective properties of LMXBs
which we will be describing in section \ref{scheme} is capable of 
handling all such prescriptions of magnetic braking. However, we
are interested here in determining the dependence of the collective 
properties of LMXBs on the very basic ingredients of the process of
formation and evolution, and not on their details. We therefore assume 
the simple form given in eqn. \ref{eqn:mb} in this work, postponing 
the study of more complicated effects to future works.

A detached pre-LMXB will shrink through the loss of angular
momentum which is caused by gravitational radiation and magnetic 
braking, described in Eq.(\ref{eqn:jorbdiff}) by
$(\dot{J}/J)_{NML} = (\dot{J}/J)_{GW} + (\dot{J}/J)_{MB}$, together
with Eqs.(\ref{eqn:gw}) and (\ref{eqn:mb}). 
We note that angular momentum loss
due to gravitational waves as well as the magnetic braking depends on the
instanteneous binary parameters, \ie, the two masses, the orbital separation
and the radius of the companion for magnetic braking. Consequently,
the evolution rate of the parameters of the system at any instant is
completely determined by their current values. This fact can be used
to evolve the system in a straightforward scheme till the point of Roche 
lobe contact, which is given by $R_c = a r_L$, where for the Roche
lobe radius $r_L$ we use the Eggleton prescription, as described in 
Paper II \citep{eggleton}.

\subsection{The bifurcation period}

Two main channels possible for LMXB evolution under  
the above conditions have been identified for a long time now.  When
the binary is sufficiently close to begin with, \ie, when the initial binary
period is shorter than the so-called \emph{bifurcation period}, 
the companion fills its Roche lobe while it is still on the main
sequence, mass transfer starts, and the LMXB turns on. Further 
orbital evolution in this channel, which is followed by close 
LMXBs and CVs, is determined by angular momentum loss, 
and the orbital period decreases during this evolution. 
On the other hand, if
the initial binary period is longer than the bifurcation period, the 
companion fills its Roche lobe after it finishes its main-sequence  
evolution and starts ascending the giant branch. In this channel,
which is followed by the wider LMXBs, further orbital evolution is 
basically determined by nuclear evolution of the companion,
and the orbital period increases during this evolution, which 
explains why it is called the bifurcation period.

Detailed calculations of this bifurcation period (which depends 
on the stellar masses and weakly on other parameters)  were given 
by \citet{pylyser88, pylyser89}, showing that this period lies 
in the range 14 -18 hours for the range of masses and parameters 
relevant for LMXBs. In the following subsections, we describe our 
calculations of LMXB evolution in the above two channels.
 
\subsection{Evolution of LMXBs with main-sequence companions}
\label{mscomp}

We consider first the evolution of LMXBs with main-sequence
companions and initial orbital periods below the bifurcation period.
Once Roche-lobe contact is established, mass transfer begins from
the low-mass main-sequence companion to the neutron star. Mass 
transferred through the $L_1$-point carries a specific angular momentum 
comparable to that of the binary orbit. Therefore, this mass cannot 
directly accrete onto the neutron star but rather forms an accretion
disk around it, from where it is slowly transferred to the neutron star on 
a viscous timescale. We in this work do not consider the detailed
dynamics of this accretion process, as it is not relevant for our
purposes. All we need for our calculations here is the fact that the 
average accretion rate onto the neutron star (\ie, averaged over any
fluctuations caused by accretion-disk processes) is
$\gamma\dot{M}$, which determines the average X-ray luminosity. 

The condition for Roche-lobe contact described in the previous 
section, \ie, $R_c = a r_L$, assumes in effect an infinitely sharp
boundary of the companion star at its formal radius, where the 
stellar density suddenly falls to zero beyond this radius. 
Such a strict condition is of course unphysical and can be
relaxed to accomodate a stellar atmosphere of rapidly falling density. We
here follow the prescription by \citet{ritter} of an exponentially 
decreasing mass transfer rate for larger separations, given by:

\begin{equation}
\dot{M} = \dot{M}_0 \; \exp\left(\frac{a r_L - R_c}{H}\right)
\label{eqn:atmos}
\end{equation}

$\dot{M}_0$ here is the rate of mass transfer exactly at the start of full
Roche-lobe overflow and $H$ is the scale height of the atmosphere. We take
$H/R_c = 0.005$ for main-sequence companions and $H/R_c = 0.01$ on the
giant branch. Note that we set the modification factor to unity in case
of a slight \emph{overfilling} of the Roche lobe, in order to avoid the very
high, umphysical rates of mass transfer which would formally result from the
above exponential prescription, and which merely reflect the fact
that this prescription is not valid when the Roche lobe is overfilled.
This method of handling the beginning of Roche-lobe overflow smoothly
is a natural way of incorporating low-luminosity systems in our
scheme. We discuss this further in Sec.\ref{properties}.

After attaining full Roche lobe contact, mass transfer begins at its
full rate from the companion to the neutron star. The response of the 
orbital separation to the mass transfer is dependent on the mass ratio $q$.
It is a well-known fact that, if there is no mass loss (\ie, $\gamma = 1$), 
then for $q < 1$ the orbit expands whereas for $q > 1$ the orbit contracts. 
Thus the mass-transfer effects strengthen those of the angular momentum loss 
for $q > 1$, while for $q < 1$ the two effects oppose each other.

We also note that, for sustained mass transfer, the radius of the star must
always equal the Roche-lobe radius. This condition is automatically
self-sustained for $q < 1$, which can be easily seen in the following way.
For Roche-lobe underfilling systems, since no mass is transferred ,
the orbit shrinks due to angular momentum loss until Roche-lobe
contact is re-established, while for Roche-lobe overfilling systems, 
heavy mass transfer occurs and the orbit widens until exact Roche-lobe
contact is re-established. For $q > 1$, the situation is more
complicated, but still viable unless $q$ has a relatively large value
(see below). For Roche-lobe underfilling systems, the same argument
as above applies. But for Roche-lobe overfilling systems, mass
transfer further shrinks the orbit and likely leads to runaway mass transfer
and heavy mass loss from the companion. It is believed that this
eventually brings the companion back into exact Roche-lobe contact.
However, at sufficiently high values of $q$, even if the entire mass
lost by the companion is lost from the system ($\gamma = 0$), the
orbit does not expand, and the above argument fails. This situation is 
similar to the one described by \citet{podsiadlowski} for companion 
masses $>4 \Msun$. Since such situations do not appear relevant for 
LMXBs, however, we do not consider them any further here, and
assume that the system is in \emph{constant}
Roche lobe contact. This condition is twofold. It implies that (a) the
radius of the companion is equal to its Roche lobe radius, and (b) the
rate of change of its radius also equals the rate of change of the
Roche lobe radius. The latter condition can be rewritten as

\begin{equation}
\frac{\dot{R}_c}{R_c} = \frac{\dot{a}}{a}\ + \frac{\dot{r}_L}{r_L}
\label{eqn:perpetrl}
\end{equation}

For a complete description of the evolution, a mass-radius relation for 
the companion is also required. Since we assume the companion to be 
on the main sequence in this channel of LMXB evolution, we can take 
a simple power-law relation $R_c \propto M_c^n$. We note that the 
mass-radius realtion for the companion could, in principle, evolve 
because of the nuclear evolution of the low-mass companion as  
the evolution of a given LMXB system proceeds, but this effect is 
unimportant for the following reason. In this channel,   
LMXB evolution occurs on the timescale of angular-momentum loss from
the system, which is shorter than the nuclear timescale of the
companion. Further, as mass transfer proceeds and the compaion's mass 
decreases, its nuclear-evolution timescale becomes longer and
longer, so that the extent of the companion's nuclear evolution 
over the entire LMXB evolution becomes quite tiny. Thus, nuclear
evolution of the companion is in effect ``frozen'', to use the 
succint description of \citet{pylyser88}, over the whole
LMXB evolution of interest to us here. Accordingly, we assume the
above mass-radius relation to be static. In the computations 
reported in this paper, we use $n=1$, which is adequate for our
first, approximate study here, which is intended to serve as a proof of
principle. Other, slightly different, values of $n$ are also discussed
later when appropriate occasions come up.

We emphasize here that the above assumption of a static mass-radius
relation would, of course, be completely inadequate for giant companions. 
We give an account of our handling of the time dependence of the 
mass-radius relation on the giant branch in Sec.\ref{giantcomp}. 
However, in this part of the work, this relation is static, so that  
the change in the radius of the mass-transferring companion is only 
through the change in its mass, and for the power-law mass-radius 
relation given above, we have $(\dot{R}_c/R_c) = -n (\dot{M}/M_c)$, 
remembering the earlier relation between $\dot{M}$ and $\dot{M_c}$. 
 
The rate of change of the Roche lobe radius can be computed by 
differentiating the Eggleton expression for the effective Roche lobe 
radius given in Paper II. We obtain
\begin{eqnarray}
\dot{r}_L &=& \frac{-\dot{M}}{M_c}\,\frac{dr_L}{dq}\,q(1+\gamma q) \nonumber \\
\frac{dr_L}{dq} &=& \frac{r_L^2}{1.47 p^4}
\left[\frac{2 \ln (1+p)}{p} - \frac{1}{1+p} \right]
\label{eqn:rlch}
\end{eqnarray}
Here, $p = q^{1/3}$. Equations (\ref{eqn:perpetrl}) and (\ref{eqn:rlch}) can be
susbstituted in Eq.(\ref{eqn:jorbdiff}) with the above mass-radius relation to 
obtain the relation between the rate of mass transfer and the rate of 
angular momentum loss. This relation is:
\begin{eqnarray}
\left(\frac{\dot{J}}{J}\right)_{NML} &=& g(q; \gamma, \mu) \frac{\dot{M}}{M_c}
\nonumber \\
g (q; \gamma, \mu) &=&
\left[ \gamma q - 1 + \frac{(1-\gamma)(1+2\mu)}{2}\frac{q}{1+q}
  -\frac{n}{2} + \frac{1+\gamma q}{2}\frac{d\ln r_L}{d\ln q} \right]
\label{eqn:jdiffms}
\end{eqnarray}

Equation(\ref{eqn:jdiffms}) gives the rate of mass transfer in terms
of the current system parameters. $\dot{M}$ and $\dot{a}$ are also 
related to each other by Eq.(\ref{eqn:perpetrl}), with the aid of the 
above mass-radius relation. Thus the evolution of a LMXB with a main 
sequence companion is completely specified by the above equations 
for given values of the parameters $\gamma$ and $\mu$. 

We take $\mu = 1$ for main-sequence companions in this part of the work.
A determination of $\gamma$ is not very straightforward, but we can
proceed in the following way. First, we shall assume in this work that 
the value of $\gamma$ is as high as possible, subject
to an absolute upper limit of unity. This assumption effectively means 
that the system's dynamics and hydrodynamics are such that it prefers 
channeling the mass through the $L_1$ point rather than sending it out 
of the system as far as possible, consistent with all laws of motion.
This is, of course, a simple assumption of convenience in a first
study, whose effects can be studied in later, detailed studies.
The second constraint comes from the fact that there is an 
upper limit to the accretion rate on the neutron star, which 
corresponds to the Eddington luminosity for the mass of the neutron
star. Finally, Eq.(\ref{eqn:jdiffms}) requires $g(q; \gamma, \mu) <
0$. This poses an additional constraint on $\gamma$, given by
\begin{equation}
\gamma \left(q - \frac{1+2\mu}{2}\frac{q}{1+q} 
+ \frac{q^2}{2r_L}\frac{dr_L}{dq}\right) < 1 + \frac{n}{2}
- \frac{1+2\mu}{2}\frac{q}{1+q}  - \frac{q}{2r_L}\frac{dr_L}{dq}
\label{eqn:glimms}
\end{equation}
We apply the above constraints to obtain a consistent value
of $\gamma$, which we use to calculate the evolution of the system
with the aid of Eq.(\ref{eqn:jdiffms}).

\begin{figure}[ht]
  \centering
  \includegraphics[scale=1.0]{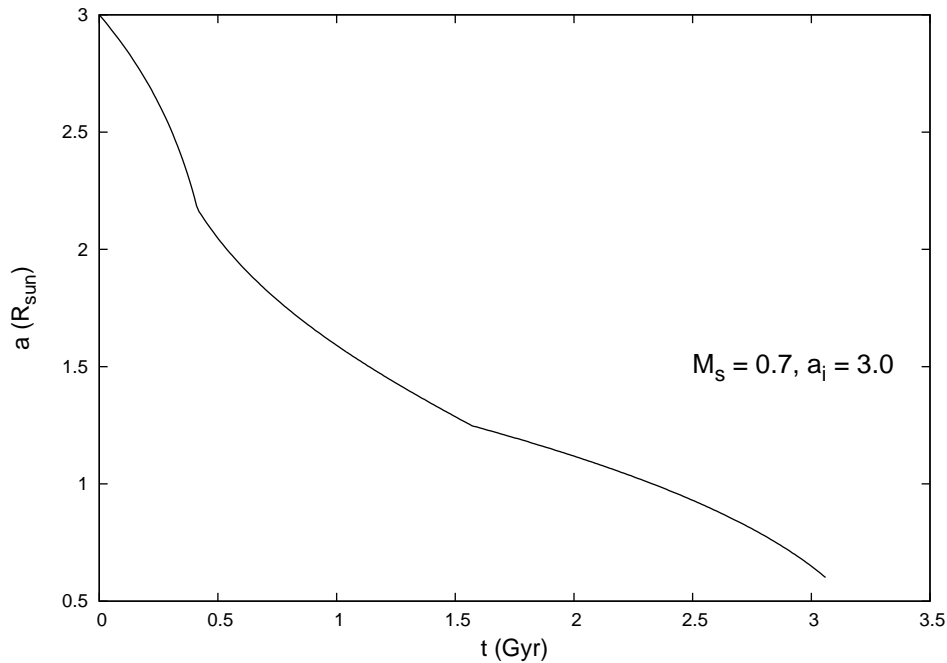}
  \vspace{1cm}
  \caption{\it Angular-momentum-loss driven orbital evolution 
           of a typical LMXB with a main-sequence companion
           (see text).}
  \label{fig:levol_ms_a}
\end{figure}

The essential features of the evolution of such LMXBs (and the related
CV systems) with low-mass main-sequence companions, where
angular-momentum loss from the system drives the mass transfer and 
the evolution, are well-known. Consider first the evolution of the
LMXB orbit, as shown in Fig.\ref{fig:levol_ms_a} for a system with an initial companion 
mass of $M_s = 0.7\Msun$ and an initial separation of $a_i = 3.0
\Rsun$, which corresponds to an initial orbital period of $P_{orb}^i 
\approx 10$ hr, \ie, shorter than the bifurcation period. 
As time progresses, the LMXB orbit shrinks 
and the orbital period decreases, even as the companion mass 
decreases due to mass transfer. Magnetic braking turns off when the 
companion becomes fully convective at a mass $M_C \approx 0.3\Msun$.
Beyond this point, the system continues evolution due to
angular-momentum loss through gravitational radiation alone, until hydrogen 
burning is extinguished in the companion below a critical mass $M_c \sim 
0.1\Msun$. Subsequently, the companion follows the mass-radius
relation for degenerate stars, the orbit expands after passing
through the so-called period minimum, and the final product is 
a close binary consisting of a recycled neutron star and a low-mass
He white dwarf. For our purposes here, it is
sufficient to follow the main-sequence phase of the companion, and 
accordingly we terminate our computations when the companion mass 
reaches $M_c = 0.1$. As a consequence, we do not follow the system
quite upto the period minimum in Fig.\ref{fig:levol_ms_a}, but otherwise the period 
evolution is very similar to that given in previous works (see,
\eg, Fig.3 of \citet{pylyser88}, Fig.3 of \citet{podsiadlowski}, 
top panel, blue curve), and the magnetic-breaking 
turn-off point shows up as a ``kink'' in this diagram (at 
$t\approx 1.5$ Gyr in Fig.\ref{fig:levol_ms_a}), as it does in the above diagram 
of \citeauthor{podsiadlowski}.

\begin{figure}[ht]
  \centering
  \includegraphics[scale=0.5, angle=270]{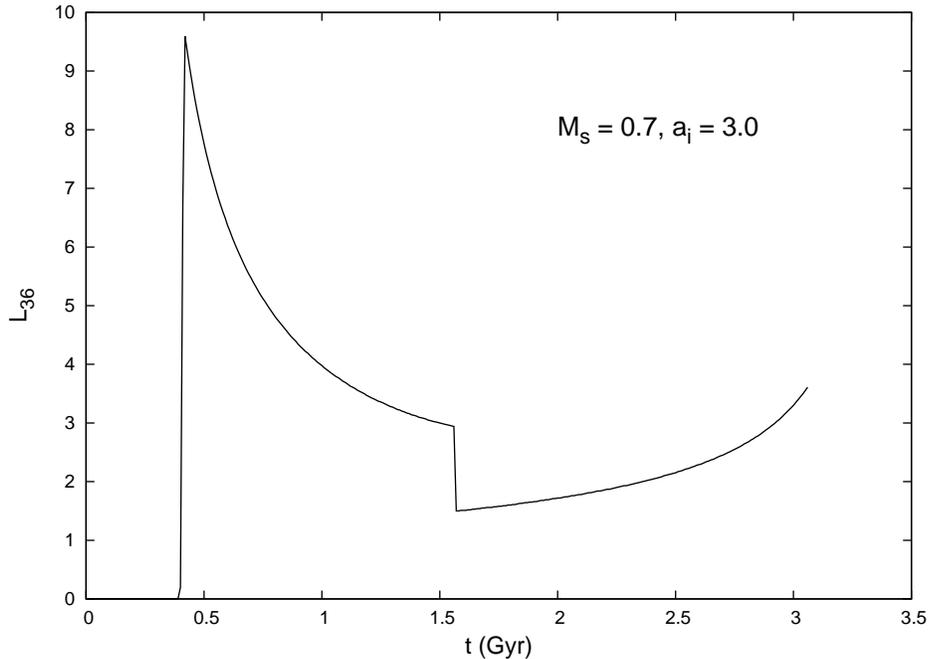}
  \vspace{1cm}
  \caption{\it Angular-momentum-loss driven luminosity evolution 
           of a typical LMXB with a main-sequence companion
           (see text and Fig.\ref{fig:levol_ms_a}).}
  \label{fig:levol_ms}
\end{figure}

Crucial for us in this work is the evolution of the accretion
luminosity of the LMXB, which we display in Fig.\ref{fig:levol_ms}   
for the system whose orbital evolution is shown in Fig.\ref{fig:levol_ms_a}.

The origin on the time axis in this figure is the formation time of 
the pre-LMXB, as before. 
The system comes into Roche-lobe contact at $t\approx 
0.4$ Gyr (which shows up as another kink in Fig.\ref{fig:levol_ms_a}), at which point
the LMXB turns on, giving the initial luminosity spike in 
Fig.\ref{fig:levol_ms}. As time progresses beyond this, there is
a sharp decline in the luminosity until the magnetic braking 
turn-off point is reached at $t\approx 1.5$ Gyr (see above), 
at which time the accretion rate and so the luminosity drops 
dramatically, since the dominant source of angular-momentum 
loss is suddenly turned off. During the subsequent evolution, 
gravitational radiation is the only mechanism of angular-momentum 
loss, and the luminosity rises very slowly with time. 

The above luminosity profile (\ie, $L-t$ relation) is in general 
agreement with those in previous works, an example of which is 
in the above \citet{podsiadlowski} work (bottom panel of 
their Fig.3, blue curve). Note that, since time is plotted on a
logarithmic scale by these authors, the initial rise in the luminosity
is actually resolved in their figure, while in our Fig.\ref{fig:levol_ms},
where we have plotted time on a linear scale, it appears as a
luminosity spike, as it does in other previous works with a linear
time plot. The basic arguments which determine the shape of this 
luminosity profile are interesting, and go back to two scaling laws 
originally demonstrated by \citet{patterson} in a related but  
different context. We briefly recount these arguments here, as their 
importance does not seem to have been fully appreciated in the LMXB 
literature. 

The $L-t$ profile, or equivalently the $\dot{M}-t$ profile, is computed
by relating the relative rate of change of the companion mass, 
$\dot{M}/M_c$, to the relative rate of change of angular momentum,
$\dot{J}/J$, as detailed in Sec.\ref{evolsec}. For clarifying the
basic scalings, we shall give a simpler calculation here, wherein we
shall use somewhat simpler prescriptions than in the detailed 
computations of Sec.\ref{evolsec}, and obtain analytic results in the
two regimes where angular momentum loss by either gravitational 
radiation (the GR regime) or magnetic braking (the MB regime)
dominates. 

We note first that in the limit $\gamma = 1$, \ie, no mass loss from
the system (which is a good approximation for the  
angular-momentum-loss driven systems with main-sequence companions 
considered in this section, as our detailed computations have shown,
as well as previous works), Eq.(\ref{eqn:jorbdiff}) reduces to:
\begin{equation}
\left(\frac{\dot{J}}{J}\right)_{NML} = \frac{\dot{M}}{M_c}
\left(q - 1\right) + \frac{\dot{a}}{2a} .
\label{eqn:jdifsimple0}
\end{equation}
In the GR regime, the left-hand side of the above equation is given
by Eq.(\ref{eqn:gw}), while in the MB regime, it is gievn by 
Eq.(\ref{eqn:mb}). The second term on the right-hand side of this 
equation is basically the relative rate of change of the orbital
radius, which can be related to the relative rate of change of the
companion mass, $\frac{\dot{M}}{M_c}$, with the aid of two 
conditions. The first is the Roche-lobe filling condition, whose
differential form is given by Eq.(\ref{eqn:perpetrl}), and the 
second is the mass-radius relation, $R_c \propto M_c^n$, whose 
differential form, $(\dot{R}_c/R_c) = -n (\dot{M}/M_c)$ was also
given earlier. When this is done, we obtain
\begin{equation}
\left(\frac{\dot{J}}{J}\right)_{NML} = \frac{\dot{M}}{M_c}
\left[-\left(\frac{n}{2} + 1\right) + \left\{q + \frac{1+q}{2}.
\frac{d\ln r_L}{d\ln q}\right\}\right],
\label{eqn:jdifsimple}
\end{equation}       
which is, of course, the simplified version of Eq.(\ref{eqn:jdiffms}) 
for $\gamma = 1$.

We now introduce a further simplification which makes the 
calculation transparent and simple for our purposes of scaling
demonstration in this part of the paper. Instead of the more
complicated, but more accurate, Eggleton prescription (see 
Sec.\ref{jloss} and Paper II), which we used throughout our
detailed computations described earlier, we now use the simpler,
original Paczynski (1971) prescription, which is still a good
approximation for the low compaion masses involved here, namely:
\begin{equation}
r_L \propto \left(\frac{q}{1+q}\right)^{1/3}
\label{eqn:paczynski}
\end{equation}     
This simplifies the calculation, yielding 
\begin{equation}
\frac{d\ln r_L}{d\ln q} = \frac{1}{3}.\frac{1}{1+q},
\label{eqn:rlsimple} 
\end{equation}
which immediately converts Eq.(\ref{eqn:jdifsimple}) into a 
very simple form
\begin{equation}
\left(\frac{\dot{J}}{J}\right)_{NML} = -\frac{\dot{M}}{M_c}
\left(\frac{5}{6} + \frac{n}{2} - q\right),
\label{eqn:jverysimple}
\end{equation}
which is useful for our purposes here. If we take $n = 1$
for lower-main-sequence stars, the quantity within the 
parentheses on the right-hand side of the above equation reduces 
to $(4/3 - q)$.

We can now obtain our analytic approximations in the GR and MB
regimes by equating the left-hand side of 
Eq.(\ref{eqn:jverysimple}) to the right-hand sides of  
Eqs.(\ref{eqn:gw}) and (\ref{eqn:mb}) respectively, and 
re-expressing the stellar masses in terms of the total mass
$M_t$ and the mass-ratio $q$ in the ensuing calculations (this
is useful because $M_t$ remains constant in these approximate
calculations due to our assumption of $\gamma = 1$ as given
above). After some straightforward algebra, we get in the GR
regime a profile:
\begin{equation}
L_{GR} \propto \dot{M}_{GR} \propto \frac{a^{\frac{10-12n}{3n-1}}}
{(1+q)\left(1-\frac{6}{5+3n}.q\right)},
\label{eqn:profilegr}
\end{equation}
where the time dependence comes from the fact that both $a$
and $q$ in the above equation are time-dependent, both 
decreasing with time, and the $a-t$ profile being given in
Fig.\ref{fig:levol_ms_a}. In the special case $n = 1$ (see above), the profile
is:
\begin{equation}
L_{GR} \propto \dot{M}_{GR} \propto \frac{a^{-1}}{(1+q)(1-0.75q)}.
\label{eqn:profilegr1}
\end{equation}            

The profile in the MB regime is obtained in a similar way, 
and found to be   
\begin{equation}
L_{MB} \propto \dot{M}_{MB} \propto a^{\frac{12n}{3n-1}}.
\frac{(1+q)^{5(n-1/3)}}{\left(1-\frac{6}{5+3n}.q\right)}.
\label{eqn:profilemb}
\end{equation} 
In the special case $n = 1$ (see above), the profile
is:
\begin{equation}
L_{MB} \propto \dot{M}_{MB} \propto a^6\frac{(1+q)^{10/3}}{(1-0.75q)}.
\label{eqn:profilemb1}
\end{equation}            

The basic physics underlying the luminosity profile in 
Fig.\ref{fig:levol_ms} is clear now from the scalings evident
in Eqs.(\ref{eqn:profilegr1}) and (\ref{eqn:profilemb1}). 
Both the GR and the MB profile factorize into $a$-dependent
and $q$-dependent parts, and the profile is largely determined
by the $a$-dependent part, as the $q$-dependent part changes
relatively slowly (compared to the $a$-dependent part) as 
the low-mass companion loses mass and $q$ decreases with 
ongoing mass transfer. Of course, since the $a$-dependence
is not as strong for GR case as it is in the MB case, some
modification does come in from the $q$-dependence, but the 
qualitative argument still holds. The sharp decline in $L$
in the MB-dominated regime between the LMXB turn-on and the 
MB turn-off at $t\approx 1.5$ Gyr is basically due to the 
very strong $a^6$-dependence of the MB luminosity in a 
shrinking orbit, as seen in Eq.(\ref{eqn:profilemb1}). 
Similarly, the slow rise in $L$ in the GR-dominated regime
beyond the MB turn-off is basically due to the 
$a^{-1}$-dependence of the GR luminosity in a shrinking 
orbit (duly modified by the $q$-dependence), as seen in 
Eq.(\ref{eqn:profilegr1}).       

These scalings were first pointed out by \citet{patterson}
in a study of CV and LMXB evolution, for the purpose of 
comparing the expected $\dot{M} - P_{orb}$ correlation with 
the data, largely on CVs. (The transformation between $a$ and
$P_{orb}$ is, of course, readily obtained through Kepler's
third law.) This amounts to looking at a ``snapshot'' at
the current epoch of a \emph{collection} of systems with various
stellar masses and at various stages of evolution, and comparing 
this data with the theoretical scaling. As \citeauthor{patterson}
noted, a reasonable account of the data was indeed given by
the above scaling, which was subsequently confirmed by  
\citet{pylyser88}, by comparing the results of their
detailed evolutionary computations with the same data.
We have argued above that the same scaling must also be
necessarily applicable to the evolution of a \emph{single} 
system as time proceeds, and have demonstrated that this
is indeed so. Accordingly, we shall call this scaling the 
\emph{Patterson scaling} throughout this work. 

We have given above a first-principles derivation of the 
Patterson scaling to emphasize its fundamental and transparent
nature, and also because some details were different in the
original \citeauthor{patterson} work, \eg, this author used a 
slightly different magnetic-braking model. However, we find
a close agreement between Patterson's result and ours, showing
the robustness of the scaling. In particular, for the specific
value of $n = 0.88$ that \citeauthor{patterson} adopted, our scaling
give $\dot{M}_{GR} \propto a^{-0.34} \propto P_{orb}^{-0.23}$, while
Patterson gives $\dot{M}_{GR} \propto P_{orb}^{-0.26}$. Similary,
in the MB regime, our scaling gives $\dot{M}_{MB} \propto a^6
\propto P_{orb}^{4.33}$, while Patterson gives $\dot{M}_{MB} 
\propto P_{orb}^{4.55}$. It is clear that the Patterson scaling
is at the heart of angular-momentum-loss driven luminosity 
evolution of LMXBs with main-sequence companions found in this 
work and numerous previous works in the subject.

\subsection{Evolution of LMXBs with giant-branch companions}
\label{giantcomp}

When the pre-LMXB is not sufficiently close according to the criterion
described earlier, \ie, the binary period is longer
than the bifurcation period, the companion finishes its main-sequence
evolution before Roche lobe contact. In such a case, the LMXB phase
turns on when the companion goes into the giant branch, starts
expanding rapidly and overfills the Roche lobe. The actual transition
timescale from the main sequence to the giant branch is that of the  
traversal of the Hertzsprung gap in the HR diagram, which is extremely
short compared to the typical timescale of LMXB evolution, and hence 
can be considered practically instanteneous for our purposes here. The
orbit cannot expand in step with this very rapid process to
accommodate the expanding giant companion, so that large mass
losses occur from the system, until the Roche-lobe filling 
condition is satisfied again. We assume in this work that there is no 
accretion during this short phase, \ie, we set $\gamma = 0$ in it.
The angular momentum loss due to the non-mass-loss mechanisms 
are also negligible due to the short duration of this
phase. Consequently, the evolution of the system can be described by 
integrating Eq.(\ref{eqn:jorbdiff}) under these assumptions, which
gives the condition:

\begin{equation}
M^i_c\sqrt{\frac{a^i}{(M^i_t)^{1+2\mu}}} = 
M^f_c\sqrt{\frac{a^f}{(M^f_t)^{1+2\mu}}}
\label{eqn:msgb}
\end{equation}

Here, the superscripts $i$ and $f$ denote the initial and the final state 
respectively. This equation is to be used along with the condition that the 
orbital separation after this phase is just enough to fit a giant 
donor within its Roche lobe, \ie,
$a^f=R_{GB}(M^f_c)/r_L(M^f_c/M_{NS})$.

In such LMXBs with giant companions, the mass-transfer rate is determined
by the rate of expansion of the companion due to its nuclear evolution. 
We assume in this work that the condition of constant Roche-lobe
filling is still valid on the giant branch and Eq.(\ref{eqn:perpetrl}) 
holds for these systems as well. However, the assumption
of a static mass-radius relation is not valid on the giant branch, and
the time evolution of the companion radius needs to be 
determined in the following way. 
The radius of a star on the giant branch depends upon its mass 
as well as its luminosity. Now, the luminosity on the giant branch 
depends upon the core mass, which, in turn, depends upon the initial 
mass rather than the current mass of the companion. Thus the evolution 
of a LMXB with a giant donor depends upon the initial values of the 
parameters \emph{as well as} their current values. With the aid of 
Eq.(48) of \citet{hurleysse}, we can write the radius a star on the giant 
branch as: 
\begin{equation}
R_c = R_{GB} = 1.1 M_c^{-0.3}(L_c^{0.4} + 0.383L_c^{0.76}).
\label{eqn:rgb}
\end{equation}
Here, $L_c$ is the instanteneous luminosity of the companion. If we define 
$f(L_c) \equiv (L_c^{0.4} + 0.383L_c^{0.76})$, rate of change of the stellar
radius can be given as
\begin{equation}
\frac{\dot{R}_c}{R_c} = 0.3 \frac{\dot{M}}{M_c} + 
\frac{1}{f} \frac{df}{dL_c} \dot{L}_c
\label{eqn:rgbch}
\end{equation}

Analytical fitting formulae for numerical stellar-evolution results 
provided by \citeauthor{hurleysse} can be used to calculate $L_c$ and 
$\dot{L}_c$, which, in turn, give the rate of expansion of the stellar radius. 
Eqs.(\ref{eqn:perpetrl}) and (\ref{eqn:rlch}) can now be inserted, along with
the new mass-radius relation given by Eqs.(\ref{eqn:rgb}) and
(\ref{eqn:rgbch}), into Eq.(\ref{eqn:jorbdiff}) to obtain the
equation governing the LMXB evolution for giant donors. We get:
\begin{eqnarray}
\left(\frac{\dot{J}}{J}\right)_{NML} &=& h(q; \gamma, \mu) \frac{\dot{M}}{M_c}
+ \frac{1}{2f} \frac{df}{dL_c} \dot{L}_c \nonumber \\
h (q; \gamma, \mu) &=&
\left[ \gamma q - 1 + \frac{(1-\gamma)(1+2\mu)}{2}\frac{q}{1+q}
  +0.15 + \frac{q(1+\gamma q)}{2 r_L}\frac{dr_L}{dq} \right] 
\label{eqn:jdiffgb}
\end{eqnarray}

Values of $\mu$ and $\gamma$ may be different here from those for
the main-sequence donor case. 
Whereas the value of $\mu$ is completely unknown as per
current understanding of the subject, it is customarily taken as
unity in studies of populations of LMXBs \citep{podsiadlowski}. 
We argue here, however, that the actual value in case of giant donors 
may well be less than this, since the outer envelope, which is the actual 
supplier of the transferred mass, is less tightly bound in the case of 
giants. We therefore adopt $\mu = 0.75$ in our calculations.

The value of $\gamma$ is subject to similar constraints as in
the case of main-sequence donors. However, we note that in case of giants
some mass loss would be inevitable. Therefore the upper limit
on the value of $\gamma$ is unlikely to be unity in the case of giant 
companions. This value is generally taken to be 0.5 \citep{podsiadlowski,
belc08}, but this is a somewhat arbitrary assumption, as noted by
\citeauthor{podsiadlowski}. We therefore keep this as a free parameter
and study its effects on the evolution of LMXB systems. Of course,
the other two constraints described earlier, namely, (a) that due to the 
Eddington limit on the mass accreting onto the neutron star, and
(b) that expressed by the condition $h(q; \gamma, \mu) < 0$, still apply.
The latter condition can be written explicitly as:
\begin{equation}
\gamma \left(q - \frac{1+2\mu}{2}\frac{q}{1+q} 
+ \frac{q^2}{2r_L}\frac{dr_L}{dq}\right) < 0.85
- \frac{1+2\mu}{2}\frac{q}{1+q}  - \frac{q}{2r_L}\frac{dr_L}{dq}.
\label{eqn:glimgb}
\end{equation}
The value of $\gamma$ is obtained subject to the three conditions
given above, and it is then used to compute the evolution of the LMXB, 
which is described by Eq.(\ref{eqn:jdiffgb}). We emphasize here
that a knowledge of the current values of the system parameters is 
\emph{not} sufficient for calculating LMXB evolution with giant-branch 
donors, which makes the computations more complicated.

LMXBs with the giant donors end their evolution when they run out of 
the supply of transferrable mass. This means that the mass transfer 
basically continues until the entire envelope of the companion is 
transferred to the neutron star, leaving only its core. We can 
write this condition as $M_c = M_{s,c}$ where $M_{s,c}$ is the core 
mass of the companion, which depends upon its initial mass. This 
He-core now contracts basically like an isolated He-star, so that
Roche-lobe contact is lost and the LMXB is turned off. We stop our 
numerical computations at this point. The final product of such
evolution is a wide, circular binary consisting of a recycled
neutron star and a low-mass white dwarf with $M_{WD} \le 0.45\Msun$
or so. 

\begin{figure}[ht]
  \centering
  \includegraphics[scale=1.0]{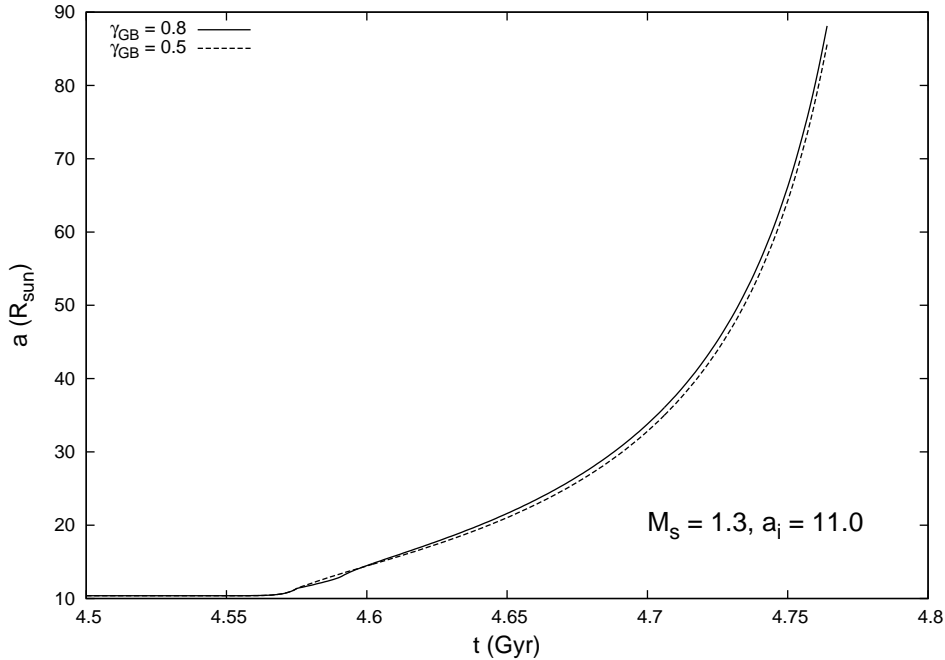}
  \caption{\it Nuclear-evolution driven orbital evolution 
           of a typical LMXB with a giant companion
           (see text). Line-style of curves coded 
           by the values of $\gamma_{max}$ as indicated.}
  \label{fig:levol_gb_a}
\end{figure}

\begin{figure}[ht]
  \centering
  \includegraphics[scale=1.0]{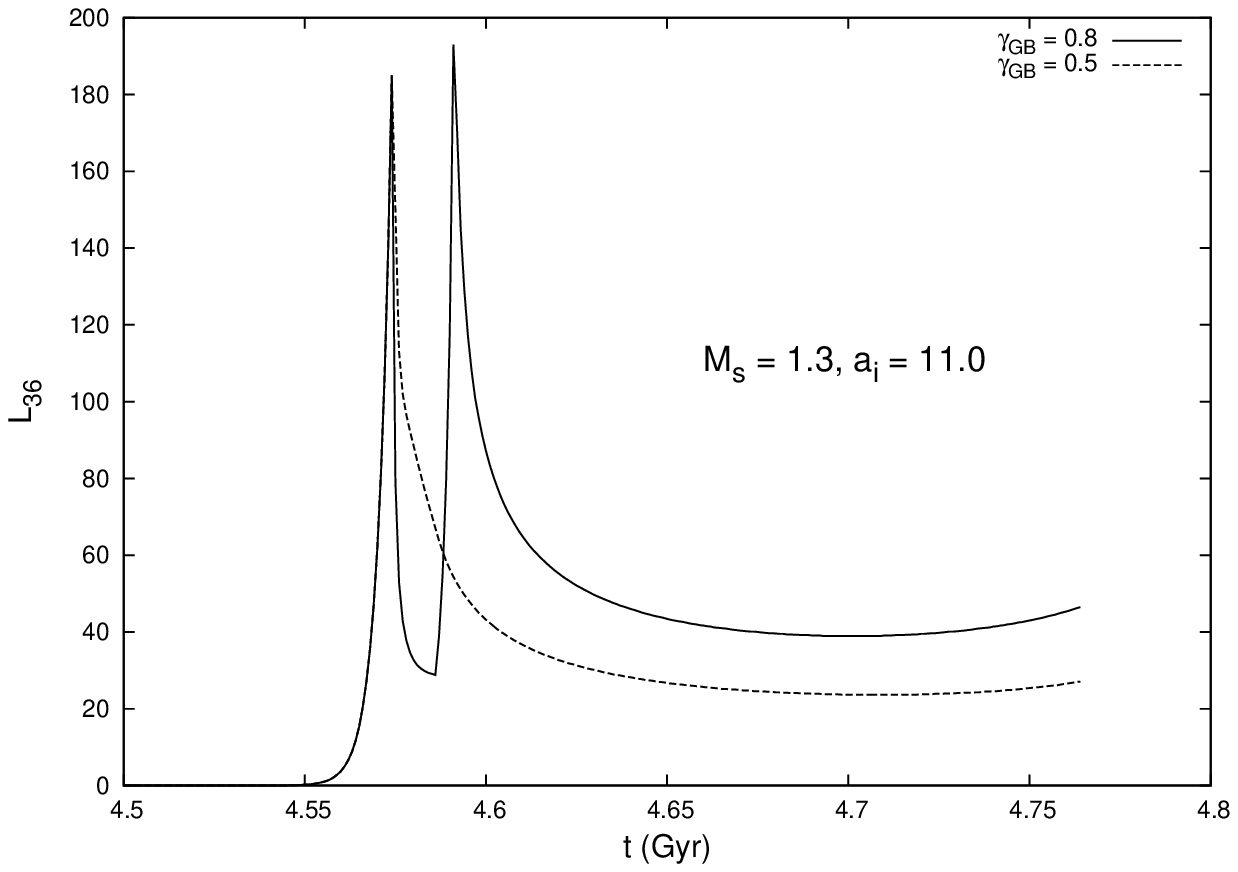}
  \caption{\it Nuclear-evolution driven luminosity evolution 
           of a typical LMXB with a giant companion
           (see text and Fig.\ref{fig:levol_gb_a}). Line-style of curves coded 
           by the values of $\gamma_{max}$ as indicated.}
  \label{fig:levol_gb}
\end{figure}

Figure \ref{fig:levol_gb} depicts the luminosity evolution of such a 
prototype system, the corresponding orbital evolution being shown in
Fig.\ref{fig:levol_gb_a}. In this case, the initial secondary mass is $M_s = 1.3$ and 
the initial orbital separation is $a_i = 11$, corresponding to an 
initial orbital period of $P_{orb}^i \approx 2.6$ days. In sharp 
contrast to the earlier angular-momentum-loss driven case, here 
the orbit \emph{expands} continually as evolution proceeds, as found 
in previous works (see, \eg, Figs.2b and 2c of \citet{tauris}, 
Fig.11 of \citet{podsiadlowski}, Figs.9 and 10 of \citet{belc08}, 
which are StarTrack studies of the above 2b and 2c cases
in the Tauris-Savonije work). In these figures, zero on the time axis 
corresponds to the formation of the pre-LMXB, as before, but since the 
system remains out of Roche-lobe contact during the secondary's main 
sequence life, we have so shifted and rescaled the time axis as to 
show most clearly only the X-ray active part with non-zero LMXB
luminosity, after the system reaches Roche-lobe contact at 
$t \approx 4.55$ Gyr. The evolutions of both the orbit and the 
luminosity found by us are rather similar to those found in previous
works, when account is taken of the fact that the display methods
in some of these previous works are different from ours, \eg,
time plotted on a logarithmic scale in \citet{podsiadlowski},
and the decreasing companion mass instead of increasing time as the
abscissa in some panels of \citet{tauris}. The major
feature of the luminosity profile after Roche-lobe contact and
LMXB turn-on is a sharp initial decline, followed by a slower
decline and a plateau phase, and a very slight rise at the longest 
times in some cases. An explanation of this luminosity profile 
involves details of stellar evolution on the giant branch, and so
is ouside the scope of this paper, contrary to the basic, simple 
explanation in terms of Patterson scaling that was possible earlier
for the angular-momentum-loss driven evolution of LMXBs with
main-sequence companions.      
  
\section{X-ray Luminosity Function (XLF) Calculation Scheme}
\label{scheme}

\subsection{The partial XLF} 
\label{partialxlf}

The importance of the cosmic star formation history (SFH) and its effect
on the pre-LMXB formation rate was noted in Paper II. It was 
explained there that the SFH over a few Gyrs previous to the current epoch 
plays a crucial role in determining the current population of LMXBs. This 
can also be seen from the two figures given in previous section depicting the
evolution of a typical LMXB system with a main-sequence or giant-branch
companion. Figures \ref{fig:levol_ms} and \ref{fig:levol_gb} show
that, while typical evolutionary timescales as counted from the
formation of the corresponding pre-LMXBs is rather similar for LMXBs
with main-sequence and giant-branch companions, typically $\sim 3-5$
Gyrs, the timescales as counted from the turn-on points of the LMXBs
are very different, typically $\sim 3$ Gyr for LMXBs with
main-sequence companions, but typically $\sim 0.2$ Gyr, \ie, an order
of magnitude shorter, for LMXBs with giant-branch companions.  
Therefore, if we observe two LMXB systems with the same luminosity in
the current epoch, their turn-on epochs may differ by upto a few
Gyrs, depending on the natures of their companions. In order to follow
the evolution of LMXB popoulations, it is therefore essential 
to take into consideration evolutionary changes in the SFR. This
problem was first considered by \citet{ghoshwhite}, who showed that 
the peak in the number of LMXBs lags appropriately behind the SFR peak. 
They considered typical timescales of LMXB evolution that are comparable
to the timescales obtained from our calculations. With the more detailed
evolutionary scenario considered in our scheme, we are able to deal
with the various timescales more accurately, considering them as  
functions of the initial parameters of the binary. This enables us
to go one step beyond the calculations of these authors, and explore 
the evolution in the distributions of the essential collective
properties of LMXBs.

The evolution of a single LMXB detailed in Sec.\ref{evolsec}
gives the complete evolutionary history of a given LMXB with a
specified initial state, which is determined by the initial mass of the
companion ($M_c^i = M_s$, see Paper II) and the initial orbital 
separation ($a^i$, for brevity we drop the superscript from here on,
and call it just $a$). In other words, the parameters of the system, 
\eg, luminosity, orbital period, instanteneous companion mass, and 
so on can be computed at any subsequent time. With this input, 
we can now calculate the X-ray luminosity function (XLF) of LMXBs 
in the following way, and we note in passing that the same method can, 
in principle, be applied to the distribution of any other collective 
property of LMXBs.

Consider first all   
pre-LMXB systems formed with the same initial parameters. The evolution
of all these systems will be similar to one another and can be 
computed with our scheme, giving the luminosity as a function of time.
This relation between luminosity and time can then be inverted, so that,
given the luminosity, we can determine the time required to attain
it. Thus, if a given LMXB with a luminosity $L$ is observed at a current time 
$t$, we know its formation time $t_f = t - t_{lb}$, where $t_{lb}$ is the
look-back time, \ie, the time taken by the system since its formation
to reach this value of luminosity. Now, the origin of time is set in 
these calculations at the formation of the pre-LMXB.
Therefore, $t_f$ is the formation time of the corresponding pre-LMXB. Thus,
the number of systems with a luminosity $L$, observed at present, is
equal to the number of pre-LMXBs formed at time $t_f$ before, as
described in Paper II. Since the formation rate can also be computed
for the given set of initial parameters, we can calculate the number of
LMXBs at each value of luminosity. By suitable binning, then, we can
numerically obtain the XLF of these systems. We note here that this 
XLF is calculated for the systems which had a specific value of $M_s$ 
and $a$ to start with. We therefore name this the \emph{partial} XLF.

\subsection{The full XLF}
\label{netxlf}

The \emph{partial} XLF, as described in the previous subsection,
can be obtained for any initial-value set ($M_s,a$). The full, or
integrated, XLF can then be obtained by integrating over $M_s$ and
$a$, or, in case of discrete data, by summing over all 
allowed values of $M_s$ and $a$. We note
here that the formation rate of pre-LMXBs given in Paper II already
includes the distribution of $M_s$ and $a$, so that one does not require
further weighting in the process of such summation/integration. 
The range of these variables relevant for the XLF calculation was also
discussed in Paper II. We simply state these ranges here, and refer to 
Paper II for further details. The range of $M_s$ is taken to be 
$[0.1, 2.5]$. We note here that this is the initial value of the 
companion mass, so that the instanteneous value of this mass 
(denoted in this work by $M_c$) will be lower than this 
at some intermediate time of evolution, after mass transfer starts. 
Systems with higher companion masses are expected to have higher 
accretion rates and lower values of $\gamma$, implying significant
mass loss in some systems. When the systems are actually observed
as LMXBs, companions more massive than $1\Msun$
are rather unlikely to be found. Nevertheless, our scheme does allow 
higher-mass companions for completeness, so that systems like Her X-1 
can also be included in the scheme.

The range of $a$ is $[1.0, 20.0]$. Very few systems wider than
this can come into Roche-lobe contact in a Hubble time. Many systems
with companion masses at the lower end of the mass-range cannot come into 
Roche lobe contact at separations $>10\Rsun$. We note here that
the systems with very low-mass companions cannot reach the giant
phase because their main sequence lifetime is longer than the Hubble
time. This lower limit on the initial companion mass can be as
high as $0.9 \Msun$, depending upon the value of Hubble constant assumed.
We note that these constraints do not need to be posed explicitly on the
systems under consideration, as our scheme of computations is so
designed as to automatically take care of such issues related to
timescales.

The XLF thus obtained depends upon (a) system parameters at the pre-LMXB
stage, which affect the results through the formation rate of these
systems, and (b) various parameters that affect the further evolution of LMXBs. 
A complete parameter study is outside the scope of this work. We therefore 
fix the values of nearly all the parameters which we list now. At the pre-LMXB
stage, we take the CE-parameter $\alpha \lambda = 1.0$, and the metallicity 
of the primary $z = 0.02$. To calculate the formation rate of the pre-LMXBs, we
assume a \emph{Madau}-profile for the SFR, which is a peak-type profile with
$z_{max} = 0.39$ and $p = 4.6$ (see \citet{blain} and also relevant references
given in Paper II for the types of SFR profile and the parameters used to 
describe them). We study the final XLF for both values of $\beta$ introduced 
in Paper II, \viz, the uniform and the falling-power-law distribution 
for the mass ratio, and also for the \emph{with kick} and \emph{without kick} 
scenarios (this latter scenario was shown in Paper II to give results
almost identical to those for ECSN-type low kicks). 
We note here that $z=0.02$ may be an incorrect 
assumption, since many LMXBs are found in early-type galaxies, where
the progenitor stars could have been metal-poor. However, we do not test 
this assumption at this stage, noting that the effect of changing the 
metallicity can be judged from the variation of the pre-LMXB PDF calculated in 
Paper II. Values of the other parameters adopted in the calculation of the 
evolution of the LMXBs are given in Secs.\ref{mscomp} and \ref{giantcomp}. 
We mention here again that we take $\mu$ and 
$\gamma_{max}$ as unity for the evolution with main-sequence companions, where 
$\gamma_{max}$ is the upper limit for $\gamma$. $\mu = 0.75$ is taken for
the evolution with companions on the giant branch, and in this case
we study two values of $\gamma_{max}$ which are 0.5 and 0.8. We
realize that these particular choices allow us to explore only a
limited section of the full parameter space. However,
our aim here is to demonstrate that the straightforward scheme described in
this work \emph{can} be used to calculate the XLF, which can be
directly compared with observations, \ie, a proof of principle. 
We thus concentrate in this work on an exploration of the relative 
importances of various physical processes on the LMXB XLF, deferring 
parameter studies to future works.

\subsection{Properties of the calculated LMXB XLF}
\label{properties}

We first note that, due to the calculational procedure adopted here,
the cumulative XLF is easier to calculate than the differential
one. Therefore, we discuss various properties of the calculated 
cumulative XLF in this subsection. We first discuss the dependence of 
the computed XLF on the pre-LMXB parameters and then that on the 
LMXB-evolution parameters. We chose two pre-LMXB parameters to
illustrate the essential points. The first is the nature of the
primordial $q$-distribution, and the second is the SN-kick scenario.
We had noted earlier that ECSN-type kicks with $\sigma = 26.5$ km/s give 
PDFs similar to the no-kick scenario, whereas ICCSN-type kicks with 
$\sigma=265$ km/s result in a different PDF. 

\begin{figure}[ht]
  \centering
  \includegraphics[scale=0.5, angle=270]{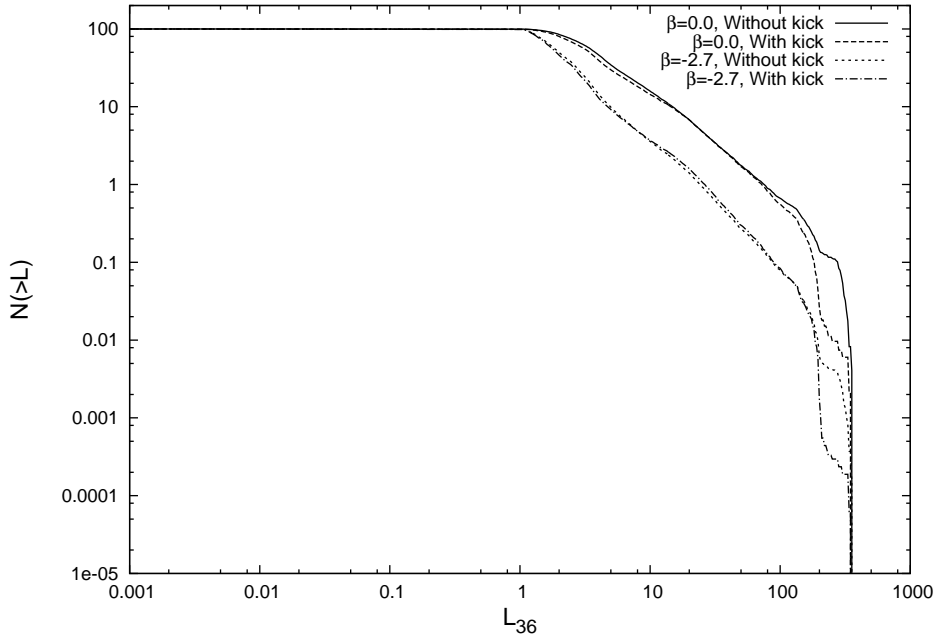}
  \vspace{1cm}
  \caption{\it Dependece of cumulative LMXB XLF on pre-LMXB
               parameters. The four cases are coded by 
               line-style according to the vaues of the parameters, 
               as indicated.}
  \label{fig:xlf_all}
\end{figure}

Fig.\ref{fig:xlf_all} shows the cumulative XLF for various values of
pre-LMXB parameters, the four cases shown being the same as those 
considered in Paper II. (For definiteness, the XLF has been normalised 
to a total of 100 systems here.) The  
following general features can be seen. The shape of the XLF is much more 
complex than that of the HMXB XLF. It starts with a flat region at low 
luminosities, which extends upto $L\approx 10^{36}$ erg/s. We discuss the 
XLF in this low-luminosity regime later in this section. A power-law like 
fall is seen at higher luminosities, which extends upto $\sim 10^{38}$ erg/s. 
The XLF cuts off beyond this point, where the luminosity approaches the 
Eddington luminosity for neutron stars. We note here that, since the mass 
of the neutron star need not be exactly the same for all systems, the 
Eddington cut-off may also have a range in general. 

It can be seen that the ``with-kick'' scenario does not produce a very different 
XLF from the ``without-kick'' scenario, except for the sharper drop near
the cutoff when the kicks are included. By contrast, the two different
$q$-distributions produce quite different XLFs in the falling-power-law
regime, with a steeper fall for the distribution of $q$ with
$\beta = -2.7$. The numerically computed XLFs in the luminosity range 
$3\times 10^{36}-10^{38}$ erg/s, as well as over other ranges, can be 
fitted by power-laws of the form $N(>L) \propto L^n$. The results of
such least-squares fits are given in Table \ref{tab1}, where the 
leftmost column shows the luminosity range considered, and the next 
columns give the values of $n$ for the four different cases. 

\begin{table}
\centering
\begin{tabular}{|c|c|c|c|c|}
\hline
Luminosity range & $\beta=0$ & $\beta=0$ & $\beta=-2.7$ & $\beta=-2.7$ \\
($\times 10^{36}$ erg/s) & without kick & with kick & without kick 
& with kick \\
\hline
3 - 10 & -1.18 & -1.16 & -1.65 & -1.51 \\
10 - 100 & -1.41 & -1.44 & -1.7 & -1.71 \\
3 - 100 & -1.31 & -1.28 & -1.59 & -1.53 \\
\hline
\end{tabular}
\caption{\it Best-fit power-law indices over different luminosity ranges
             for our computed LMXB XLF, for different values of pre-LMXB 
             parameters as indicated.}
\label{tab1}
\end{table} 

Relative contributions of main-sequence and giant-branch donors
can influence the shape of the overall LMXB XLF, since many 
characteristics of these two donor classes differ from each other. 
First, the giants come into Roche-lobe
contact later than the main-sequence companions. Thus the giant population
corresponds to an earlier population of primordial binaries. Since the
Madau profile suggests a peak in SFR at $z = z_{max} = 0.39$, this effect
can be important. Second, systems with giant companions are typically
brighter, but have shorter lifetimes. This makes them much smaller in number,
but situated at the high-$L$ end of the XLF. We study the effect of the 
fraction of systems with giant companions, which we name the 
\emph{giant fraction}, on the XLF in the following way. 
We first evolve the LMXB systems in the way
described in Sec.\ref{evolsec}. We calculate the XLF obtained from this
evolutionary scheme which includes main sequence as well as giant donors.
We then modify our algorithm of evolution in such a way that, when a 
companion completes its main sequence life, the system is removed from the 
computations. This eliminates the possibility of having giant donors,
keeping in our computation only those systems which reach
Roche-lobe contact during their main-sequence life, and so ensuring
that all LMXBs in our computation have \emph{only} main-sequence companions.  
The XLF computed in such a way is thus that corresponding to
main-sequence donors alone. 

\begin{figure}[ht]
  \centering
  \includegraphics[scale=0.5, angle=270]{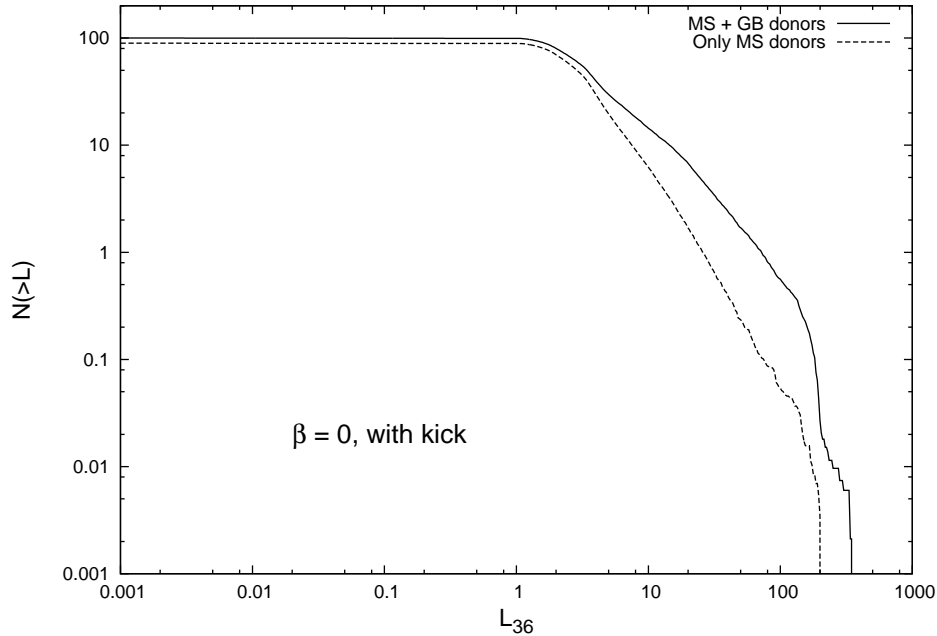}
  \vspace{1cm}
  \caption{\it Relative contributions of main-sequence and giant donors in
           the computed XLF. Cases encoded by line-style as indicated.}
  \label{fig:msgb}
\end{figure}

Figure \ref{fig:msgb} shows the XLF for only main-sequence donors, 
superposed on the XLF after the inclusion of the giants. 
We see that the total number of systems with giant companions is not
more than a few percent of the total number, but the giant-companion 
systems all lie at the bright end of the XLF, making the XLF 
considerably flatter at this end (see Table 1) than that with 
main-sequence companions alone, which has a slope of $n \approx -1.94$
at this end. We note here that, though the computation presented here
to illustrate this point is for $\beta=0$ and the ``with-kick'' case, 
the qualitative result is the same for all other combinations of 
$\beta$-value and kick-scenario. Therefore the giant fraction,
although numerically small, is a very important factor in deciding 
the shape of the LMXB XLF at the bright end.

\begin{figure}[ht]
  \centering
  \includegraphics[scale=0.5, angle=270]{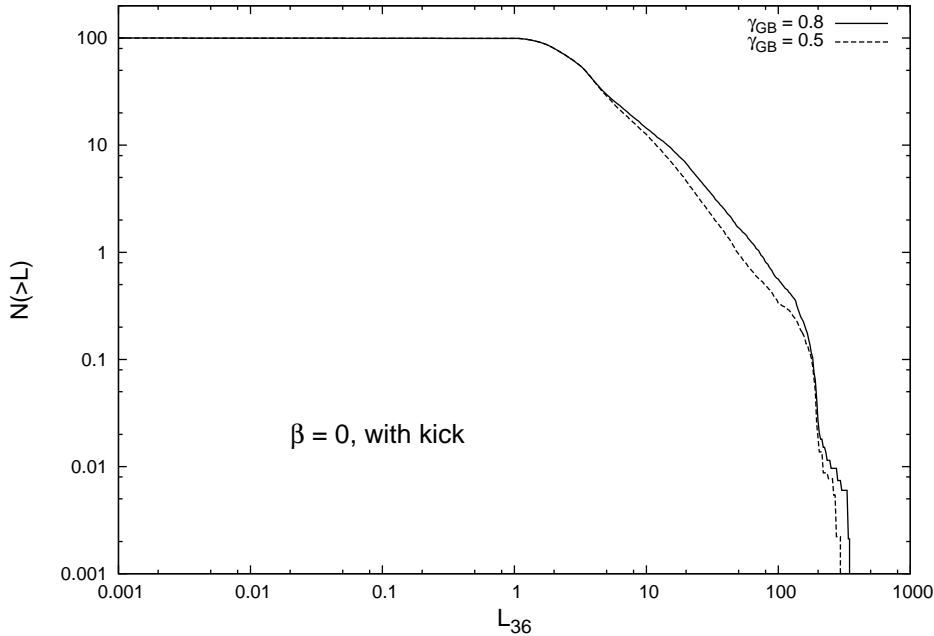}
  \vspace{1cm}
  \caption{Dependence of LMXB XLF on $\gamma_{GB}$. Curves labeled
           by the respective values of $\gamma_{GB}$.}
  \label{fig:gch}
\end{figure}

The parameter $\gamma$ denotes the accretion rate onto the neutron
star per unit rate of mass loss by the companion. This parameter is 
close to unity for main-sequence companions, but its value is expected
to be lower for giant companions, and this value was taken to be 
$\gamma_{GB}=0.8$ in previous calculations. A change in the
value of $\gamma_{GB}$ results in a change of the luminosity of those
systems with giant donors which are operating at a minimum mass loss from 
the system (see Sec.\ref{giantcomp}). The importance of the giant fraction 
in determining the shape of the XLF has already been shown above, from
which we expect that lowering the value of $\gamma_{GB}$ will lower
the influence of the giant-companion systems on the total XLF (since
the maximum allowed luminosity for these objects will be lowered),
and so make the XLF fall more steeply at the bright end.  
Figure \ref{fig:gch} shows that this is indeed the case, displaying
the effect for two values of $\gamma_{GB}$. The first value is 
0.8, which has been used throughout this work. The second value is 0.5,
which has been suggested by some authors, as noted in Sec.\ref{giantcomp}.
The power-law index obtained by least-squares fit method in the luminosity 
range $L_{36} = 3 - 100$ is $n=-1.48$ in the second case, \ie, steeper than
in the first case (see Table \ref{tab1}), but not as steep as in the
$\gamma_{GB}=0.8$ case (see above).
 
Finally, we consider the nature of our computed XLF at very low 
luminosities ($L < 10^{36}$ erg/s). The XLF at these luminosities
appears to be not very interesting for several reasons. 
First, it is difficult to observe such faint systems with the current 
sensitivity of the X-ray detectors, even in nearby galaxies. Thus the 
observed XLF will be strongly influenced by selection effects in this
region. Although studies often attempt to account for incompleteness, 
there are necessarily large error-bars in observed XLFs in this regime,
as a study of relevant works shows. Second, one can see from our 
calculations that the standard theory predicts that there would be
very few systems in this range. In fact, if we exclude Roche-lobe 
underfilling companions, the minimum luminosity possible is 
$\approx 10^{36}$ erg/s, which corresponds to a binary of a 1.4\Msun\ 
neutron star and a main-sequence companion with a mass just below 
0.3\Msun, the critical point where the magentic braking is turned off
(corresponding roughly to the minimum luminosity in 
Fig.\ref{fig:levol_ms}). Only
with our inclusion of Roche-lobe underfilling companions and the
Ritter recipe for atmospheric Roche-lobe overflow (wherein the 
accretion rate drops exponentially for increasing disparity between 
the stellar radius and the Roche-lobe radius; see
Eq.(\ref{eqn:atmos})), as detailed in Sec.\ref{mscomp} do we account 
for systems at lower luminosity.  

\begin{figure}[ht]
  \centering
  \includegraphics[scale=0.5, angle=270]{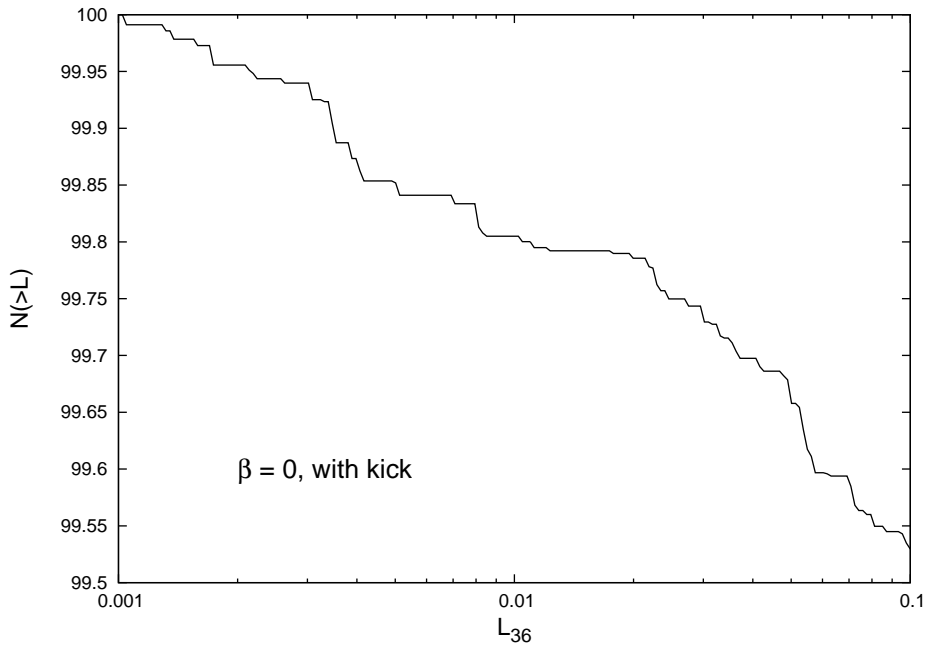}
  \vspace{1cm}
  \caption{\it LMXB XLF at very low luminosities.}
  \label{fig:faint}
\end{figure}

Figure \ref{fig:faint} shows the details of the computed XLF in this 
luminosity regime, using the above Ritter recipe. 
Because of the very narrow range of the
ordinate ($N(>L)$) involved in this regime, this part of the XLF
appeared almost flat (\ie, constant $N(>L)$) in all previous XLFs
shown in this paper, where the scale of $N(>L)$ was logarithmic.
In this figure, we have displayed $N(>L)$ on a linear scale, and
shown only that range of $N(>L)$ over which it varies in this regime,
in order to show the nature of its variation. Apart from a artifacts
resulting from binning and numerical effects, the average trend is
consistent with a logarithmic decrease in $N(>L)$ with increasing
$L$-values.

\section{Discussion}
\label{discuss}

\subsection{Comparison with observations}
\label{observe}

In recent years, LMXB XLFs have been obtained for many early-type
galaxies from \emph{Chandra} and \emph{XMM-Newton} observations. 
Many such works have come to the conclusion that the observed LMXB
XLF cannot be described by a simple, single-power-law distribution, 
unlike the HMXB XLF \citep{grimm01, humphrey, kim, revnivtsev08, 
voss06, voss09, fragos09}. These works have argued that the LMXB XLF 
can be adequately described by a broken power-law with a high luminosity 
cut-off. It was suggested by \citet{grimm01} that the number of LMXBs 
in a galaxy would scale with the \emph{total stellar mass} of that
galaxy, so that, when the LMXB XLF is normalised by this stellar mass, 
a roughly \emph{universal} LMXF XLF would emerge, similar to the 
situation for the HMXB XLF and its scaling with the \emph{current SFR}
in the particular galaxy, which was described in Paper I. However, 
this scaling was not found to be as precise as that in the case of HMXBs. 

It was shown by \citet{gilfanov04} that a doubly-broken power-law
(\ie, a power law with \emph{two} breaks) template gave an adequate fit to
the observed universal LMXB XLF. The break points obtained by these 
authors were at $L_{36}=19$ and $L_{36}=500$, and the power-law 
exponents given by them for the \emph{differential} XLF $dN/dL$ 
in the three regions separated by these breaks were $-1.0 \pm 0.13, 
-1.86 \pm 0.12$ and $-4.8 \pm 1.1$, going from low to high 
luminosities. We stress again that these authors provided 
fits to the \emph{differential} XLF, \ie, $dN/dL$. Power-law indices
in this description are obtained by subtracting $1$ from the ones in the 
cumulative description, except in the special case of a power-law
index of $-1$ in the differential description, where in the corresponding 
cumulative description $N(>L)$ is a logarithmically decreasing
function of $L$.

However, \citet{kim} proposed a different template. These 
authors showed that the XLF of a \emph{single} galaxy can be described
adequately by a single power-law. With a sample of 14 E and S0 galaxies,
these authors concluded that  power-law exponents in the range $(-0.8,
-1.2)$ could explain the \emph{cumulative } XLFs of all galaxies
in their sample, and proposed further that a single power-law with an 
exponent $-1.1 \pm 0.1$ for the cumulative XLF would be consistent
with the combined data. However, these authors also pointed out that 
a broken power-law would provide an improved fit. The break-point 
suggested by these authors roughly agrees with the second 
break-point of \citeauthor{gilfanov04}. The power-law slopes 
given by \citet{kim} for the \emph{differential} XLF were $-0.8 \pm 0.2$ 
and $-1.8 \pm 0.6$ respectively, below and above this break. Thus, a broken 
power-law (with one or two breaks) generally seems to account for all
observations of LMXB XLFs to date, and we shall use this as the template
for our discussion here. 

However, comparing these observed XLFs with our computed XLF 
is not completely straightforward, due to the complex nature of the
XLFs. Note first that observed XLFs include NS as well BH binaries, and
so extend upto luminosities $\sim 10^{39}$ erg/s. Systems beyond the second
break in the above template are clearly BH systems, whereas their
contributions  below the second break, which need not be zero, are
unknown. By contrast, the calculations presented by us in Paper II 
and in this work include only NS systems, so that a strong cut-off
is expected and seen at the neutron-star Eddington luminosity.
The power-law regime in our computed XLF is in the range $L_{36}=(3, 100)$,
which roughly overlaps with the middle region between the two breaks 
in the observed ``universal'' XLF, which is $L_{36}=(20, 500)$. We
note from Table \ref{tab1} that our $\beta=0$ case gives a power-law
slope relatively close to that of the observed XLF of
\citet{gilfanov04}, whereas the $\beta=-2.7$ case gives an XLF
considerably steeper than the observed one. Our results for the 
$\beta=0$ case also match reasonably with the results of \citet{kim} 
for the single power-law fit to individual galaxies.

On the whole, it appears at this stage of comparison between observed
and computed XLFs that $\beta=0$ provides a better match than
$\beta=-2.7$. However, it remains to be explored if different
appropriate choices of other parameters might not give a power-law 
slope in better agreement with observations even for the latter 
value of $\beta$.

\subsection{Conclusions}
\label{conclude}

We remind ourselves that, while the computed XLF presented in this
paper depends closely on the computed pre-LMXB PDFs presented
in Paper II, pre-LMXBs are, by and large, not observable. Accordingly,
the LMXB XLF comparisons presented in this paper (and possibly such 
comparisons of calculable distributions of other observable collective 
properties of LMXBs) serve as a major check on the above  pre-LMXB 
PDFs. To the extent that even a qualitative account of observed 
LMXB XLFs is possible by our simple scheme, a certain measure of
confidence in the inferred pre-LMXB PDFs of Paper II is gained from
these investigations, which has implications for studies of related
types of X-ray binaries.

We saw in the previous section that the giant fraction is an important 
parameter in determining the shape of the XLF at the bright end. 
We emphasize that this 
fraction is \emph{not} a free parameter in our calculational scheme, 
set by hand. Rather, it is decided by the shape of the pre-LMXB PDF.
Thus, a pre-LMXB PDF which gives a larger giant fraction produces a 
shallower XLF. At one remove, this corresponds to the inclusion of a 
higher fraction of wider pre-LMXBs with higher-mass companions. 
The choices made by us for primordial-binary parameters and subsequent
evloutionary parameters in Paper II determined this fraction entirely,
leaving no free parameters to adjust at the stage of computing the
XLF. 

A primordial-binary parameter which seems to be very important for
its ultimate influence on the XLF shape is that which controls the 
shape of the $q$-distribution, \viz, the parameter $\beta$. The 
reason for this can be traced to the influence of $\beta$ on the
pre-LMXB PDF for both $M_s$ and $a$, as shown in Paper II
(see Figs. 6 and 11 of that paper). Between the $q$-distribution and the 
SN-kick scenario, the former has by far the dominant influence
on the pre-LMXB PDF, when other parameters are held fixed. It
comes as no surprise, therefore, that $\beta$ should have a 
strong influence on the XLF shape. What we do find  
interesting is that, of the two values of $\beta$ used in previous works
on the subject, one fares so much better than the other by the XLF
criterion. However, we must remind ourselves that this has been 
shown to be true in this work only for the chosen values of the 
other parameters. It would be premature to take the results of this
work as a definite indication that a flat $q$-distribution in
primordial-binary distribution is always preferred, until the 
phase-space of the other parameters is thoroughly studied.
However, to the extent that the values of the latter parameters are
representative of actual pre-LMXBs, our result does seem 
suggestive.     

By contrast, the SN-kick scenario has little influence on the XLF,
as expected form the above arguments, since it has little effect on
the pre-LMXB PDF, as shown in Paper II. Indeed, the only case
in which it has any significant effect is that of the $a$-distribution
in the $\beta=0$ case (see Fig.11 of Paper II). But even this has little
final influence on the XLF, since the $M_s$-distribution is much
more effective in controlling the XLF than the $a$-distribution.

The calculations of various rates of the mass transfer and the orbital 
shrinking with our prescription of the mass loss depend crucially upon the 
assumption of \emph{constant} Roche lobe contact during mass transfer
and loss (if any). We have given arguments in favor of this assumption
at appropriate places in this work. However, we stress here that this 
assumption should be made with appropriate timescales in mind, as we
now explain. The typical timescales of the LMXB evolution are $\sim
2-3$ Gyrs. A timestep of $10^{-2}$ Gyrs is therefore sufficiently short for
keeping an accurate track of the evolution of these systems, and we
have, of course, verified that finer time-steps give essentially the
same result. This is tantamount to assuming that the orbit readjusts
itself to Roche-lobe contact within the timescale of the step size 
assumed, \ie, $\sim 10^7$ yr, which appears to be a safe assumption. 

With the cautionary remark that, since the falling-power-law region in 
our calculated XLF does not exactly overlap with the power-law region 
suggested by empirical fits to the observations, we should be careful
about drawing conclusions, it does seem significant that a uniform 
primordial $q$-distribution does consistently lead to a better agreement
with observations than the falling-power-law $q$-distribution with a 
slope of -2.7. With due caution, we therefore suggest that our results
do indicate that a $\beta = 0$ distribution for LMXB-progenitor
primordial binaries seems to be favored by the XLF observations. 
It would be interesting to see if the inclusion of black-hole sytems 
leads to an exact match with the observations, which would be a rather
gratifying confirmation of our suggestion for the underlying
primordial $q$-distribution. We note, however, that the other parameters
held fixed in this study, \eg, the CE parameter, may also   
have significant roles in determining the slope in this region. The value
of $\gamma_{max}$ on the giant branch is also uncertain. However, low
values of this parameter around 0.5 would decrease the effect of the giant 
populations and hence are less likely. It may also be not impossible
that this parameter gradually decreases with time as the companion
star evolves along the giant branch. As we saw earlier, the value  
$\gamma_{max}$ = 0.8 produces a better fit with the data. This may
thus represent an average value of the $\gamma_{max}$ over this entire 
evolution.

The problems of a proper representation of the LMXB XLF at the 
lowest-luminosity end are many, not the least of which is the
possibility that the accretion paradigm in this region may be entirely
different, \eg, ADAF (see \citet{narayan08, lasota08} for recent
reviews) or related flow models. Incorporation of such models into
our scheme is a very ambitious task, since the relations between the
mass-transfer rate, the mass-accretion rate, and the output X-ray
luminosity are thought to be very different in some of these models
from the simple accretion-disk paradigm we have used here. Accordingly,
we defer the inclusions of both black-hole systems and such acretion
flow models to future works. We conclude by re-emphasizing that our
effort in this series of papers must be regarded as an attempt at
a proof-of-principles exploration of whether observed collective
properties of X-ray binaries can be accounted for by evolving 
canonical collective properties of primordial bianries through 
well-accepted scenarios of individual binary evolution. Considering
the simplicity of our scheme (for both HMXBs and LMXBs), the
agreement for HMXBs was remarkable, and that for LMXBs, although
not as precise, is still qualitatively in the correct direction.
Thus encouraged, we feel justified in attempting more elaborate 
future explorations, which would incorporate detailed 
stellar-evolution models.

\bibliographystyle{apj}
\bibliography{bibfile}

\begin{thebibliography}{27}
\expandafter\ifx\csname natexlab\endcsname\relax\def\natexlab#1{#1}\fi

\bibitem[{{Belczynski} {et~al.}(2008){Belczynski}, {Kalogera}, {Rasio}, {Taam},
  {Zezas}, {Bulik}, {Maccarone}, \& {Ivanova}}]{belc08}
{Belczynski}, K., {Kalogera}, V., {Rasio}, F.~A., {Taam}, R.~E., {Zezas}, A.,
  {Bulik}, T., {Maccarone}, T.~J., \& {Ivanova}, N. 2008, \apjs, 174, 223

\bibitem[{{Blain} {et~al.}(1999){Blain}, {Smail}, {Ivison}, \& {Kneib}}]{blain}
{Blain}, A.~W., {Smail}, I., {Ivison}, R.~J., \& {Kneib}, J.-P. 1999, \mnras,
  302, 632

\bibitem[{{Eggleton}(1983)}]{eggleton}
{Eggleton}, P.~P. 1983, \apj, 268, 368

\bibitem[{{Faulkner}(1971)}]{faulkner}
{Faulkner}, J. 1971, \apjl, 170, L99+

\bibitem[{{Fragos} {et~al.}(2009){Fragos}, {Kalogera}, {Willems}, {Belczynski},
  {Fabbiano}, {Brassington}, {Kim}, {Angelini}, {Davies}, {Gallagher}, {King},
  {Pellegrini}, {Trinchieri}, {Zepf}, \& {Zezas}}]{fragos09}
{Fragos}, T., {et~al.} 2009, \apjl, 702, L143

\bibitem[{{Ghosh} \& {White}(2001)}]{ghoshwhite}
{Ghosh}, P., \& {White}, N.~E. 2001, \apjl, 559, L97

\bibitem[{{Gilfanov}(2004)}]{gilfanov04}
{Gilfanov}, M. 2004, \mnras, 349, 146

\bibitem[{{Grimm} {et~al.}(2002){Grimm}, {Gilfanov}, \& {Sunyaev}}]{grimm01}
{Grimm}, H.-J., {Gilfanov}, M., \& {Sunyaev}, R. 2002, \aap, 391, 923

\bibitem[{{Humphrey} \& {Buote}(2004)}]{humphrey}
{Humphrey}, P.~J., \& {Buote}, D.~A. 2004, \apj, 612, 848

\bibitem[{{Hurley} {et~al.}(2000){Hurley}, {Pols}, \& {Tout}}]{hurleysse}
{Hurley}, J.~R., {Pols}, O.~R., \& {Tout}, C.~A. 2000, \mnras, 315, 543

\bibitem[{{Kim} \& {Fabbiano}(2004)}]{kim}
{Kim}, D.-W., \& {Fabbiano}, G. 2004, \apj, 611, 846

\bibitem[{{Lasota}(2008)}]{lasota08}
{Lasota}, J.-P. 2008, \nar, 51, 752

\bibitem[{{Narayan} \& {McClintock}(2008)}]{narayan08}
{Narayan}, R., \& {McClintock}, J.~E. 2008, \nar, 51, 733

\bibitem[{{Patterson}(1984)}]{patterson}
{Patterson}, J. 1984, \apjs, 54, 443

\bibitem[{{Peters}(1964)}]{peters65}
{Peters}, P.~C. 1964, Physical Review, 136, 1224

\bibitem[{{Peters} \& {Mathews}(1963)}]{peters63}
{Peters}, P.~C., \& {Mathews}, J. 1963, Physical Review, 131, 435

\bibitem[{{Pfahl} {et~al.}(2003){Pfahl}, {Rappaport}, \&
  {Podsiadlowski}}]{pfahl03}
{Pfahl}, E., {Rappaport}, S., \& {Podsiadlowski}, P. 2003, \apj, 597, 1036

\bibitem[{{Podsiadlowski} {et~al.}(2002){Podsiadlowski}, {Rappaport}, \&
  {Pfahl}}]{podsiadlowski}
{Podsiadlowski}, P., {Rappaport}, S., \& {Pfahl}, E.~D. 2002, \apj, 565, 1107

\bibitem[{{Pylyser} \& {Savonije}(1988)}]{pylyser88}
{Pylyser}, E., \& {Savonije}, G.~J. 1988, \aap, 191, 57

\bibitem[{{Pylyser} \& {Savonije}(1989)}]{pylyser89}
{Pylyser}, E.~H.~P., \& {Savonije}, G.~J. 1989, \aap, 208, 52

\bibitem[{{Revnivtsev} {et~al.}(2008){Revnivtsev}, {Lutovinov}, {Churazov},
  {Sazonov}, {Gilfanov}, {Grebenev}, \& {Sunyaev}}]{revnivtsev08}
{Revnivtsev}, M., {Lutovinov}, A., {Churazov}, E., {Sazonov}, S., {Gilfanov},
  M., {Grebenev}, S., \& {Sunyaev}, R. 2008, \aap, 491, 209

\bibitem[{{Ritter}(1988)}]{ritter}
{Ritter}, H. 1988, \aap, 202, 93

\bibitem[{{Tauris} \& {Savonije}(1999)}]{tauris}
{Tauris}, T.~M., \& {Savonije}, G.~J. 1999, \aap, 350, 928

\bibitem[{{van der Sluys} {et~al.}(2005){van der Sluys}, {Verbunt}, \&
  {Pols}}]{sluys}
{van der Sluys}, M.~V., {Verbunt}, F., \& {Pols}, O.~R. 2005, \aap, 440, 973

\bibitem[{{Verbunt} \& {Zwaan}(1981)}]{verbunt}
{Verbunt}, F., \& {Zwaan}, C. 1981, \aap, 100, L7

\bibitem[{{Voss} \& {Gilfanov}(2006)}]{voss06}
{Voss}, R., \& {Gilfanov}, M. 2006, \aap, 447, 71

\bibitem[{{Voss} {et~al.}(2009){Voss}, {Gilfanov}, {Sivakoff}, {Kraft},
  {Jord{\'a}n}, {Raychaudhury}, {Birkinshaw}, {Brassington}, {Croston},
  {Evans}, {Forman}, {Hardcastle}, {Harris}, {Jones}, {Juett}, {Murray},
  {Sarazin}, {Woodley}, \& {Worrall}}]{voss09}
{Voss}, R., {et~al.} 2009, \apj, 701, 471

\end{thebibliography}

\end{document}